\documentclass[aps,pre,amsmath,amsfonts,amssymb,twocolumn,showpacs]{revtex4}
\usepackage{bm}
\usepackage{epsfig}
\newcommand{\x}{{\bm{x}}}
\newcommand{\y}{{\bm{y}}}
\newcommand{\e}{{\bm{e}}}
\newcommand{\act}{{\tt act}}
\newcommand{\lis}{{\tt list}}
\newcommand{\brh}{{\bm{\rho}}}
\begin{document}
\title{Monte Carlo simulations of bosonic reaction-diffusion systems}
\author{Su-Chan Park}
\email{psc@kias.re.kr}
\affiliation{Korea Institute for Advanced Study, Seoul 130-722, Korea}
\date{\today}
\begin{abstract}
An efficient Monte Carlo simulation method for bosonic 
reaction-diffusion systems
which are mainly used in the renormalization group (RG) study is proposed.
Using this method, one-dimensional bosonic single species annihilation model
is studied and, in turn, the results are compared with RG calculations.
The numerical data are consistent with RG predictions.
As a second application, a bosonic variant of 
the pair contact process with diffusion  (PCPD)
is simulated and shown to share the critical behavior with
the PCPD. The invariance under the Galilean transformation of this
boson model is also checked and discussion about the 
invariance in conjunction with other models are in order.
\end{abstract}
\pacs{64.60.Ht, 05.10.Ln, 89.75.Da}
\maketitle
\section{Introduction}
The reaction-diffusion (RD) systems have become a paradigm for studying
certain physical, chemical, and biological systems \cite{P97}.
In the study of the RD systems on a lattice via Monte Carlo (MC) simulations,
particles involved in the dynamics usually have hard core exclusion property.
In other words, MC simulations have been interested in 
the lattice systems where multiple occupancy at a lattice point is prohibited. 
These particles are often referred to as {\it fermions}, but
this paper prefers the term ``hard core particles.''
Meantime, the renormalization-group (RG) calculations that 
have been applied successfully to several RD systems
are in many cases based on the path integral formalism for
classical particles without hard core exclusion, or, if we
are allowed to abuse terminology, {\it bosons} \cite{bosonF,Lee94,CT96}.
On this account, the comparison of the numerical studies 
to the RG calculations is sometimes nontrivial.

There are two ways to fill a gap between numerical and analytical studies.
One is to make a path integral formula for hard core particles 
which is suitable for the RG calculations.  Actually, this path has been 
sought and some formalisms are suggested \cite{PKP00,vW01,PjP05}.
The other is to find a numerical method to simulate boson systems.
In this context, numerical integration studies of the equivalent 
Langevin equations to boson systems have been performed 
\cite{BAF97,CD99,PL99,DCM04}.
However, it is not always possible to find an equivalent Langevin equation
\cite{G83}. By the same token, the applicability of this approach is 
somewhat restricted. Thus, another numerical method is called for.
To our knowledge, no algorithm to simulate general bosonic RD systems directly  
has been suggested and to find such a algorithm is still a challenging topic.

This paper suggests an algorithm to simulate the bosonic RD systems.
Section \ref{Sec:algo} is devoted to a heuristic explanation of 
the algorithm to simulate general bosonic single species RD systems.
In Sec. \ref{Sec:app}, the numerical method applies to some
bosonic RD systems. At first, the single species annihilation
models with various conditions are simulated, 
along with the comparison to the  RG predictions.
Then, a bosonic version of the pair contact process with diffusion 
is discussed, focusing on the universality and Galilean invariance.
Section \ref{Sec:sum} summarizes the works.

\section{\label{Sec:algo}Algorithm}
This section explains the algorithm suitable for MC simulations of
bosonic RD systems.
Although the discussion in this section is restricted to single species cases, 
the extension to multispecies problems is straightforward.

The reaction dynamics of diffusing bosons is represented as
\begin{equation}
n A \stackrel{\lambda_{nm}}{\longrightarrow}  (n+m) A,
\label{Eq:dynamics}
\end{equation}
where $n\ge 0$, $m \ge -n$, $m\neq 0$,
 and $\lambda_{nm}$ is the transition rate.
Each particle diffuses with rate $D$ on a 
$d$ dimensional hypercubic lattice.
The periodic-boundary conditions are assumed, but 
other boundary conditions do not limit the validity of the algorithm.
Configurations are specified by the occupation number 
$\rho_\x$ ($\ge 0$) at each lattice point $\x$. 
A configuration is denoted as $\{ \rho \}$
which means $\{\rho_{\x} | \x \in \bm{L}^d\}$,
where $\bm{L}^d$ stands for the set of the lattice points 
and the cardinality
of $\bm{L}^d$ is $L^d$.  The master equation which describes 
stochastic processes modeled by Eq. (\ref{Eq:dynamics}) takes the form
\cite{G83,K97}
\begin{equation}
\begin{aligned}
&\frac{\partial P }{\partial t} = D \sum_{\langle \x ,\y \rangle} 
\left ( (\rho_\x + 1)  \hat E_{\x,\y} - \rho_\x \right )P\\
&+ \sum_{n,m} \lambda_{nm} \sum_{\x} \left ( f(\rho_\x-m,n) \hat C_{\x,m} 
 - f(\rho_\x ,n)  \right )P,
\end{aligned}
\label{Eq:master}
\end{equation}
where $P=P(\{\rho\},t)$ is the probability with which the configuration
of the system is $\{\rho\}$ at time $t$, 
$\langle \x,\y \rangle$ means the nearest-neighbor pair ($\x,\y \in \bm{L}^d$), 
$f(\rho_\x,n) = (\rho_\x !)/(\rho_\x -n )!$ is the number of
ordered $n$ tuples at site $\x$ of the configuration $\{\rho\}$,
and $\hat E_{\x,\y}$ and $\hat C_{\x,m}$ are operators affecting
$P(\{\rho\},t)$ such that
\begin{equation}
\begin{aligned}
\hat E_{\x,\y} P  &= P(\{\cdots,\rho_\x +1,
\rho_\y-1,\cdots\};t),  \\
\hat C_{\x,m}P  &= P(\{\cdots,\rho_\x-m,\cdots\};t). 
\end{aligned}
\end{equation}

The master equation implies that during infinitesimal time interval $dt$, 
the average number of transition events for the configuration $\{\rho\}$ is
\begin{equation}
\label{Eq:ave}
\begin{aligned}
E(dt,&\{\rho\})=
dt \sum_{\x,n} \left ( 2 d D\delta_{n,1} + \sum_m \lambda_{nm} 
\right ) f(\rho_\x,n)\\
&=dt \sum_{\x,n} \left ( 2 d D\delta_{n,1} +  \sum_m n!\lambda_{nm} 
\right ) g(\rho_\x,n),
\end{aligned}
\end{equation}
where $g(\rho_\x,n) = f(\rho_\x,n)/n! = \binom{\rho_\x}{n}$ is the 
number of (nonordered) $n$ tuples at site $\x$.
Therefore, the first step for MC simulations is
to select one of $n$ tuples with an equal probability.
For the convenience of description and better understanding, we introduce a model dependent
function $h(\rho_\x,n)
= \epsilon_n g(\rho_\x,n)$, where
$\epsilon_n$ takes 1 (0) if
$D\delta_{n,1} + \sum_m \lambda_{nm}$ is nonzero (zero).
The meaning of $\epsilon_n$ is straightforward; we do not have to consider
the reaction dynamics with transition rate zero (see below). 

\begin{figure}[t]
\includegraphics[width=0.45\textwidth]{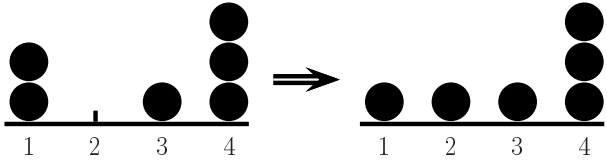}
\caption{\label{Fig:example} An example of a configuration change 
of a one-dimensional RD system with $L = 4$ due to hopping.
The black circle signifies a particle and 
numbers below the horizontal line indicate the lattice point.
A particle at  site 1 hops to  site 2.
}
\end{figure}
The simplest way to implement the selection is as follows:
First a site $\x$ is selected with probability
$N_\x / M$, where $N_\x = {\sum_n} h(\rho_\x,n)$ which 
will be called the number of accessible states at site $\x$ and 
$M = \sum_\x N_\x$. Then, $n$ is chosen with probability
$h(\rho_\x,n)/N_\x$ which is zero if $\epsilon_n=0$. 
For this procedure, the array of the number of particles at all sites,
say $\brh[~]$ ($\brh[\x] = \rho_\x$), is  necessary.

However, it is not efficient because there are too many floating number 
calculations.  
For a faster performance, we introduce two more 
arrays, say \lis[~] and \act[~][~]. 
The array \lis[~] refers the location of any $n$ tuple.
Each element of \lis[~] takes the form $(\x,\ell)$, 
where $\x$ is a site index and $\ell$ lies between 1
and $N_\x$ ($1 \le \ell \le N_\x$).
From $\ell$ and  $\brh[\x]$, 
which $n$ tuple is referred by the array \lis[~] is determined.
If $\ell \le h(\rho_\x,0)$, $n=0$  is implied. 
Else if $\ell \le h(\rho_\x,0)+ h(\rho_\x,1)$, $n=1$ is meant. 
Else if $\ell \le h(\rho_\x,0)+h(\rho_\x,1) + h(\rho_\x,2)$, 
$\ell$ indicates one of pairs at site $\x$, and so on. 
In case the total number of accessible states in the system is $M$, 
the size of \lis[~] is $M$ 
and all elements of \lis[~] should satisfy that 
$\lis[p] \neq \lis[q]$ if $p\neq q$ ($1\le p,q \le M$).
Hence, the random selection of an integer between 1 and $M$ 
is equivalent to the choice of one $n$ tuple among $M$ accessible states
with an equal probability. 
The array \act[~][~] is the inverse of the \lis[~], that is, 
$\lis[s]=(\x,\ell)$ corresponds to $\act[\x][\ell]=s$. 

\begin{table}[b]
\caption{\label{Table:example} An example of making two arrays
referring each other from the configuration shown in Fig. 
\ref{Fig:example}. Two columns on the left (right) hand side
correspond to the configuration before (after) the hopping event.}
\begin{ruledtabular}
\begin{tabular}{llll}
\multicolumn{2}{c}{Before}&\multicolumn{2}{c}{After}\\
\hline
\lis[1]=(1,1)&\act[1][1]=1&\lis[1]=(1,1)&\act[1][1]=1\\
\lis[2]=(1,2)&\act[1][2]=2&\lis[2]=(4,6)&\act[2][1]=9\\
\lis[3]=(1,3)&\act[1][3]=3&\lis[3]=(4,5)&\act[3][1]=4\\
\lis[4]=(3,1)&\act[3][1]=4&\lis[4]=(3,1)&\act[4][1]=5\\
\lis[5]=(4,1)&\act[4][1]=5&\lis[5]=(4,1)&\act[4][2]=6\\
\lis[6]=(4,2)&\act[4][2]=6&\lis[6]=(4,2)&\act[4][3]=7\\
\lis[7]=(4,3)&\act[4][3]=7&\lis[7]=(4,3)&\act[4][4]=8\\
\lis[8]=(4,4)&\act[4][4]=8&\lis[8]=(4,4)&\act[4][5]=3\\
\lis[9]=(4,5)&\act[4][5]=9&\lis[9]=(2,1)&\act[4][6]=2\\
\lis[10]=(4,6)&\act[4][6]=10&&
\end{tabular}
\end{ruledtabular}
\end{table}

After selecting $\x$ and $n$, the transition $nA \rightarrow (n+ m)A$
occurs with the probability of $n! \lambda_{nm} \Delta t$ for all possible
$m$, where $\Delta t$ is independent from configurations.
Provided $n=1$ is selected, in addition to reaction processes,
a particle at $\x$ hops to one of the
nearest neighbors with probability $D\Delta t$.
To make the transition probability have a meaning, $\Delta t$ should satisfy
\begin{equation}
\left (2 d D \delta_{n,1} + \sum_m n! \lambda_{nm} \right ) \Delta t \le 1,
\end{equation}
for all $n$. Time is increased by $\Delta t / M$.
On average, this algorithm generates $E(\Delta t,\{\rho\})$ 
transition events during time interval $\Delta t$.
After the system's evolving, three arrays, $\brh$, \lis, and \act, are
updated in a suitable way (see below).

\begin{figure}[t]
\includegraphics[width=0.37\textwidth]{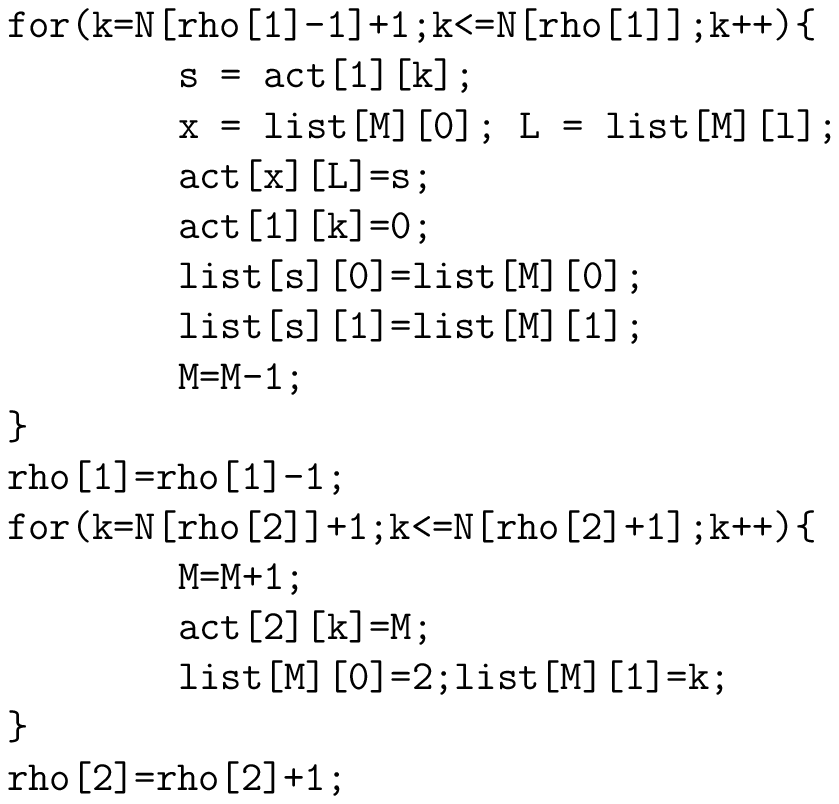}
\caption{\label{Fig:code} A program which updates three arrays 
$\brh[~]$, $\lis[~]$, and $\act[~][~]$ after the hopping 
event shown in Fig. \ref{Fig:example}.}
\end{figure}

Through an example, how the system evolves {\it in silico} is to be clarified. 
Consider a RD system with $\epsilon_n = 0$ for
$n \ge 3 \text{ and } n=0$.
In this case,  $N_\x = \sum_n h(\rho_x,n) = g(\rho_\x,1) + g(\rho_\x,2) 
= \rho_\x(\rho_\x + 1)/2$ will be used.  
Assume that we are given a  configuration
$\brh[1]=2$, $\brh[2]=0$, $\brh[3]=1$, and $\brh[4]=3$ 
($N_1=3$, $N_2=0$, $N_3 = 1$, $N_4 = 6$, hence $M=10$);
see Fig. \ref{Fig:example}.
Complete lists of two arrays \lis[~] and \act[~][~] for this configuration are 
illustrated on the left-hand side of Table \ref{Table:example}.
The algorithm starts from selecting one number between 1 and $M$, randomly.
Let us assume that 2 is selected, which makes $\lis[2]$ to be checked.
Since $\lis[2]=(1,2)$ and $2 \le \brh[1]$, a particle dynamics at  site 1 
will be attempted. Again assume that  a hopping to the site 2 whose probability
is $D\Delta t$ occurs,
which results in a change of the configuration as shown 
in Fig. \ref{Fig:example}.
Accordingly, three arrays should be updated.
Figure \ref{Fig:code} shows how the evolution is coded 
(based on the language {\sc c}).
In this code, {\tt rho}[$x$] is the number of particles at site $x$ 
($=\rho_x$), {\tt N}[$x$] is the number of accessible states at 
site $x$ ($= N_x$), and each element of \lis[~] is treated as an array.
The first (second) {\tt for} loop signifies the decreasing 
(increasing) of the number of accessible states at site 1 (2),
which can be used for any 
particle number decreasing (increasing) events.
The code generates
the lists on the right-hand side of Table \ref{Table:example}.
Time is increased by $\Delta t/10$. 
Then again choose one number between 1 to 9, randomly, and so on.

Equipped with the numerical methods, Sec. \ref{Sec:app} studies
some bosonic RD systems which show scaling behavior.
\section{\label{Sec:app}Applications}
\subsection{\label{sec:AA}Single species annihilation model}
The algorithm explained in the previous section is applied to a one-dimensional 
single species annihilation model which corresponds to $\lambda_{nm}=0$
unless $n=2$ and $m=-2$.  For saving the writing effort, 
let us rename $\lambda_{2,-2} \mapsto \lambda$.
The renormalization-group calculation predicts that 
the annihilation fixed point corresponds to $\lambda = \infty$ \cite{Lee94}.
Infinite pair annihilation rate means that two particles occupying
the same site by any chance will be removed instantaneously. Accordingly,
at most one particle can reside at each site.
Hence, the boson model with infinite annihilation rate is equivalent
to the diffusion-limited annihilation model (DLAn) of hard core particles
which can be solved exactly \cite{S00}. 
It is known that the particle density of the DLAn starting from
the random initial condition decays as 
\begin{equation}
\rho(t) = \lim_{L\rightarrow \infty} \frac{1}{L} \sum_{x=1}^L \rho_x(t) = \frac{1}{\sqrt{8 \pi D t}}( 1 + O(1/t)).
\label{Eq:AA_decay}
\end{equation}
This behavior does not depend on the initial density. 
Since renormalized coupling constant 
flows to the annihilation fixed point,
the asymptotic behavior of the density for finite $\lambda$ is
expected to be the same as Eq. (\ref{Eq:AA_decay}).
Besides, it is expected that 
the smaller the value of $\lambda$,
the later the system enters the scaling regime.
Actually, these predictions are tested for the annihilation model
of hard core particles \cite{PPK01}.
However, to our knowledge, there is no satisfactory numerical test 
for the RG predictions using a boson model \cite{comment}.

\begin{figure}[t]
\includegraphics[width=0.4\textwidth]{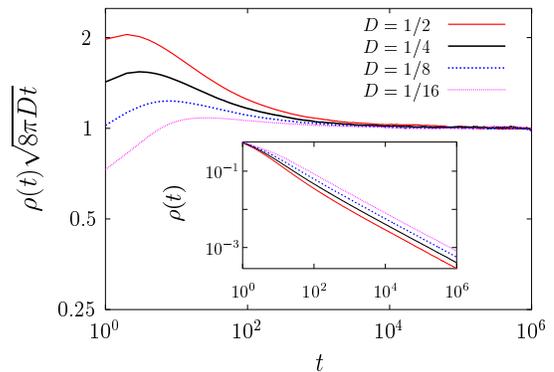}
\caption{\label{Fig:var_D} 
(Color online) A log-log plot of $\rho(t) \sqrt{8\pi D t}$ as a 
function of $t$ for various $D$ with $\lambda=\frac{1}{2}$ and $\rho_0=1$.
All curves approach to 1 as $t$ increases. Inset: same, but density
is not multiplied by $\sqrt{8\pi D t}$.}
\end{figure}
\begin{figure}[b]
\includegraphics[width=0.4\textwidth]{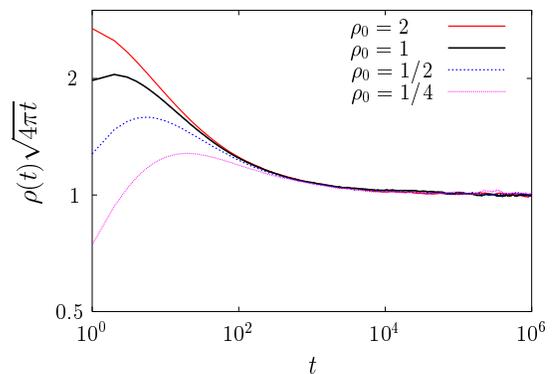}
\caption{\label{Fig:var_rho} 
(Color online)
A log-log plot of $\rho(t) \sqrt{4\pi t}$ vs $t$ for various $\rho_0$
with $D=\frac{1}{2}$ and $\lambda = \frac{1}{2}$. In the asymptotic regime, all data
sets show the same behavior.}
\end{figure}
The Poisson distribution is used as an initial condition,
which can be implemented if we randomly distributed $\rho_0 L$ particles 
on the lattice.  For this distribution, the probability that $q$ 
particles reside at site $\x$ is 
\begin{equation}
P_x(q) = \left ( \begin{array}{c} \rho_0 L \\ q \end{array} \right )
\left (\frac{1}{L} \right )^{q} \left ( 1 - \frac{1}{L} 
\right )^{\rho_0 L - q} \sim \frac{\rho_0^q}{q!}e^{-\rho_0},
\end{equation}
where $L$ is assumed to be sufficiently large and
$q \ll \rho_0 L$.
Using the algorithm explained in the previous section and varying
$D$, $\lambda$, and $\rho_0$,  we simulated the one-dimensional
annihilation model. The system size is $2^{16}$ 
and the number of independent samples is 200 for each data set.

\begin{figure}[t]
\includegraphics[width=0.4\textwidth]{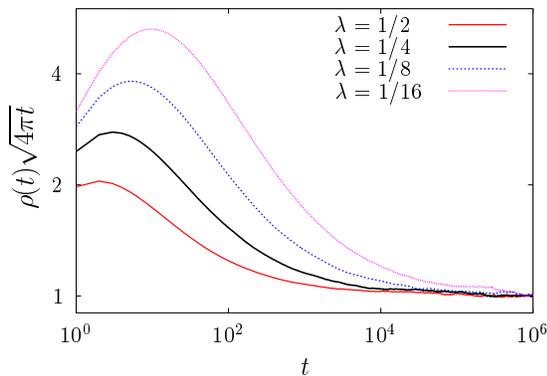}
\caption{\label{Fig:var_l} 
(Color online)
A log-log plot of $\rho(t) \sqrt{4\pi t}$ vs $t$ for various $\lambda$
with $D=\frac{1}{2}$ and $\rho_0 = 1$. 
Although the system with smaller $\lambda$
enters the scaling regime slowly, all curves eventually meet for large $t$.}
\end{figure}
Figure \ref{Fig:var_D} shows the decaying behavior of the density
for $D=\frac{1}{2}$, $\frac{1}{4}$, $\frac{1}{8}$, and 
$\frac{1}{16}$ with $\rho_0=1$ and $\lambda=\frac{1}{2}$.
Each curves approaches to $1/\sqrt{8 \pi D t}$ as the RG calculation predicted.
We also check the initial condition dependence,
by simulating systems with various initial density  2, 1, $\frac{1}{2}$, 
and $\frac{1}{4}$ with $D=\frac{1}{2}$ and $\lambda = \frac{1}{2}$.
Figure \ref{Fig:var_rho} shows  the initial condition independence 
of the asymptotic behavior.
Finally, we also confirm that the asymptotic behavior is not affected
by $\lambda$, see Fig. \ref{Fig:var_l}.
As expected, the system with smaller $\lambda$ enters the scaling regime later.
The MC simulation for bosonic annihilation models confirms
the predictions of the RG study \cite{Lee94}.
\subsection{\label{sec:pcpd}Pair contact process with diffusion}
The pair contact process with diffusion (PCPD) is a RD system of 
diffusing hard core particles with two 
competing dynamics of $2A \rightarrow 3A$ (fission) 
and $2A \rightarrow 0$ (annihilation), which
shows a continuous transition \cite{HH04}.
At first sight, the bosonic variant of the PCPD might be regarded
as the boson model with $\lambda_{nm} = 0$ except $\lambda_{21}$ 
and $\lambda_{2,-2}$. However, this variant does not show a
continuous transition and there is no steady state in its 
active (fission dominating) phase \cite{HT97}.
To have a well-defined steady state in all phases, a mechanism to keep
the density from blowing up is required. Introducing a triple 
reaction such as $3A\rightarrow 2A$, one can get a model
with well-defined steady states. Although the boson model with
$\lambda_{nm}=0$ except $\lambda_{21}$, $\lambda_{2,-2}$,
and $\lambda_{3,-1}$ has been expected to show a continuous transition
\cite{HH04}, MC simulation results 
for this type of boson model which will be called
``BPCPD'' has yet been reported in the literature,
although a parallel update bosonic model with so-called soft-constraint 
was studied \cite{KC03}.

\begin{figure}[t]
\includegraphics[width=0.45\textwidth]{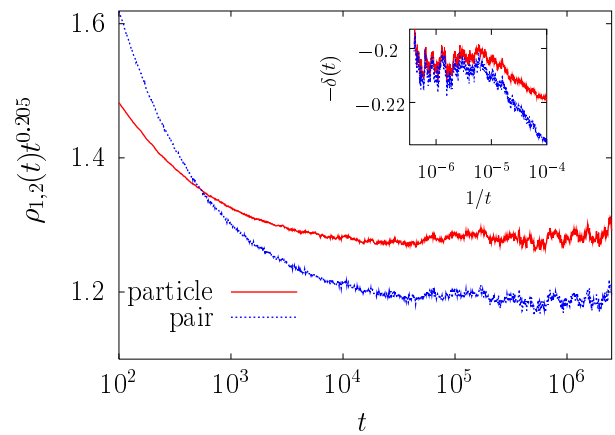}
\caption{\label{Fig:pcpd} (Color online) 
Time dependence of the particle and pair
densities multiplied by  $t^{\beta/\nu_\|}$ with $\beta/\nu_\| = 0.205$
in semilogarithmic plot at criticality for the BPCPD.  
Inset: effective exponents of the order parameters at $p=0.148~79$.}
\end{figure}

Using parameter values $D=\frac{1}{2}$, $\lambda_{3,-1} = \frac{1}{6}$, 
$\lambda_{2,-2} = p/2$, and $\lambda_{21} = (1-p)/2$ where $p$ is the 
tuning parameter, the critical behavior of the BPCPD
is studied. As an initial condition,
we set $\rho_\x = 2$ for all $\x$ ($1\le \x \le L$).
Figure \ref{Fig:pcpd} shows the decaying behavior at criticality 
of two order parameters, the particle and pair densities which are defined as
\begin{equation}
\begin{aligned}
\rho_1(t) &= \frac{1}{L} \sum_\x \langle \rho_\x \rangle_t,\\
\rho_2(t) &= \frac{1}{L} \sum_\x \langle \rho_\x (\rho_\x -1)\rangle_t,
\end{aligned}
\end{equation}
where $\langle \cdots \rangle_t$ means the average over ensembles
at time $t$.
The system size in use is $2^{15}$ and all samples (around $10^3$ samples 
are independently simulated) up to observation time ($\sim
2.5\times 10^6$ MC steps) have at least one site with two or more particles.
The critical point is found to be $p_c = 0.148~79(1)$ with the critical
exponent $\beta/\nu_\| = 0.205(5)$ which is estimated 
from the effective exponent
\begin{equation}
-\delta(t) = \frac{\ln[\rho_{1,2}(t)] - \ln[\rho_{1,2}(t/m)]}{\ln m},
\end{equation}
with $m=10$.
At criticality, $\delta(t)$ approaches to $\beta/\nu_\|$ as $t$ 
goes to infinity.
The simulation results  are consistent 
with the previous works within error bars \cite{KC03,PhP05a}.
Hence, we conclude that the BPCPD 
has the same critical scaling with the PCPD.

Following the path integral formalism for bosonic RD systems \cite{bosonF}, 
the action of the BPCPD, $S = \int dt~d{\bm x} {\cal L}$, after taking the
(naive) space-time continuum limit has the form 
\begin{equation}
{\cal L} = \bar \phi (\partial_t - D \nabla^2 ) \phi 
+ g_1 \bar \phi \phi^2  + \lambda_3 \bar \phi \phi^3
+ g_2 \bar \phi^2 \phi^2+ \cdots,
\label{Eq:pcpd_action}
\end{equation}
which is the same as one studied in Ref. \cite{JWDT04} which is
derived from path integral formalism for the exclusive particle systems
introduced in Ref. \cite{vW01}.
It is argued, however, via  RG calculations \cite{JWDT04} 
and numerical studies \cite{PhP05a,PhP05b} 
that Eq.~\eqref{Eq:pcpd_action} is 
inappropriate for studying the critical behavior of the PCPD using
the RG techniques.  Nonetheless, we will show 
that the Galilean invariance (GI) of the BPCPD, which is anticipated
from Eq.~\eqref{Eq:pcpd_action}, is still correct in the strong sense
(see below).

\begin{figure}[t]
\includegraphics[width=0.48\textwidth]{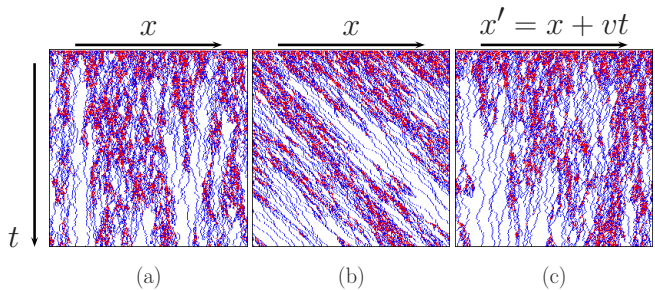}
\caption{\label{Fig:XT_config}(Color online)
Space-time configuration for the unbiased 
and biased BPCPD models at criticality. 
Blue (red) dots represent the sites with only one particle 
(at least two particles) and white dots stand for the empty sites.
(a) and (b) are configurations of the BPCPD with $D_R = \frac{1}{2}$ 
and $D_R = 1$,
respectively. (c) is the same as (b) except that the space coordinate 
is Galilean transformed with $v=D_R - D_L = 1$.}
\end{figure}
For some RD systems, biased diffusion only changes nonuniversal constants
such as the critical point and does not affect the critical behavior.
Examples are the driven branching annihilating random walks (DBAW) studied
in Ref. \cite{PhP05a}. Such systems will be called to
have the GI {\it in the weak sense} (GIweak).
Why the critical point is dependent on the bias strength is understandable
within the framework of Ref. \cite{PjP05}.
Using the path integral formalism for hard core particles introduced
in Ref. \cite{PjP05}, the terms appearing in the action due to 
the bias with the strength $v$ take the form
\begin{equation}
{\cal L}_{\textrm{bias}} = 
v \left (\bar \phi_\x   \partial_\|  \phi_\x  -  \bar \phi_\x^2 \phi_\x 
\partial_\| \phi_\x \right ),
\label{Eq:biased}
\end{equation}
where $\partial_\|$ is the lattice gradient defined as
$\partial_\| \phi_\x \equiv (\phi_{\x + \bm{e}_\|} - \phi_{\x - \bm{e}_\|} )/2$
with $\bm{e}_\|$ the unit vector along the bias direction.
The derivation of Eq. \eqref{Eq:biased} is shown in
the Appendix.
The Galilean transformation gauges away the first term in 
Eq.~\eqref{Eq:biased}, but cannot remove the second term.
Since the second term in Eq.~\eqref{Eq:biased} is irrelevant in the 
RG sense for the DBAW,
this does not affect the universal behavior, but the very existence of this
irrelevant term can change the critical point.
Therefore, the DBAW is of the GIweak.
Meanwhile, the PCPD is not of the GI even in the weak sense
\cite{PhP05a}.
Since it is shown that the field theory with the action
(\ref{Eq:pcpd_action}) is not viable \cite{JWDT04}, we cannot 
extract any information from Eq. (\ref{Eq:biased}) concerning the 
driven pair contact process with diffusion (DPCPD).
To understand the DPCPD and the PCPD from the field theoretical
point of view, more elaborated studies are required.

\begin{figure}[t]
\includegraphics[width=0.48\textwidth]{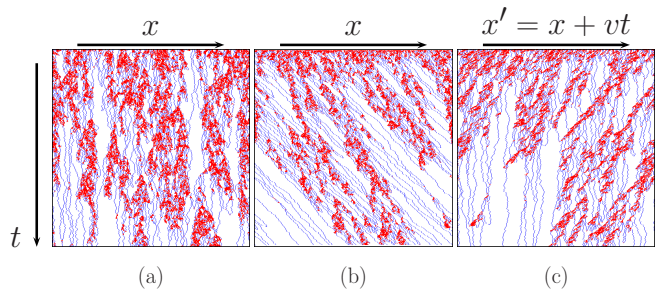}
\caption{\label{Fig:XT_config_HC}(Color online)
Space-time configuration for (a) the PCPD 
and (b) the DPCPD at criticality studied in Ref. \cite{PhP05a}.
Blue (red) dots represent the isolated particles 
(particles which are members of pairs) and white dots stand for the empty sites.
(c) is the same as (b) except that  the space coordinate 
is Galilean transformed with $v= 1$.}
\end{figure}

The bias diffusion of bosons does not generate the
second term in Eq.~\eqref{Eq:biased}.
In this context, the Galilean transformation totally gets rid of the 
effect of bias for bosons. Hence, two systems with or without bias have 
the same probability distribution, let alone the critical behavior.
These systems will be called to have the GI {\it
in the strong sense} (GIstrong).
Consider a one-dimensional bosonic RD system
with reaction dynamics in Eq. (\ref{Eq:dynamics})
in which each particle hops to the right (left) with rate
$D_R$ ($D_L$). The GIstrong for this model
means that whatever value $D_R$ takes with the constraint
$D_R + D_L = 2 D$ (constant), the system shares the probability
distribution with the unbiased model ($D_R = D_L = D$).
It is checked numerically for various $D_R$ 
with $D_R + D_L = 1$, whether the BPCPD has the GIstrong or not.
We observed that the particle 
and pair densities have the same behavior at the same $p$ 
within statistical error (not shown here).
In Fig. \ref{Fig:XT_config}, space-time configurations of the
BPCPD models with the unbiased diffusion
($D_R = D = \frac{1}{2}$), fully biased diffusion ($D_R = 2 D = 1$), and 
the Galilean transformation for the full bias case are shown.
After Galilean transformation, no noticeable difference between 
biased and unbiased cases is observed.
For comparison, we present in Fig. \ref{Fig:XT_config_HC}
the space-time configuration of the PCPD
and the DPCPD studied in Ref. \cite{PhP05a}.
As the Galilean transformed space-time configuration shows, the bias 
cannot be removed in the DPCPD. The Galilean transformation generates
the biased motion of the paired particles which shows
the existence of the relative bias between isolated particles and
paired ones.
Although the validity of Eq. (\ref{Eq:pcpd_action}) 
as an appropriate action for the RG study regarding the PCPD is rather 
problematic, any single species bosonic RD systems with on-site reactions are
conjectured to have the GIstrong.

The discussion about the GIstrong should be
restricted to boson models with random sequential update dynamics.
If the dynamics occurs in a parallel way as in Ref. \cite{KC03}, 
the GI argument from the invariance of the local action like
Eq. (\ref{Eq:pcpd_action})
under the Galilean transformation is not directly applicable.
Even worse, the one dimensional system with $p_R=1$ 
(for the definition of $p_R$, see
the next paragraph) is reduced to a single-site problem
which is not expected to show phase transition.
Notwithstanding, except this pathological case, 
the soft-constraint PCPD (SCPCPD) studied in Ref. \cite{KC03}
is expected to have the GIweak \cite{KCcomment}.

To understand what is happening in the SCPCPD, let us explain
the dynamics of the model.
During unit time, changes of a configuration occur in two steps.
At first, every particle hops to the right (left) with probability
$p_R$ ($p_L$) and stays still with probability $p_S$ 
($p_R + p_L + p_S = 1$). 
In Ref. \cite{KC03}, $p_R=p_L=\frac{1}{2}$ and
$p_S = 0$ are used. After the hopping events, reactions occur
at all sites.
Rather interestingly, the model with $p_L=0$ is statistically 
equivalent to the system with $p_S = 0$ provided $p_R$ is the same.
When $p_S=0$, particles at the even sites do not interact with 
those in the odd sites. For example, see  Fig. 
\ref{Fig:example} and regard the left figure of it as a configuration for 
the SCPCPD with $p_S=0$ under the condition of the periodic boundary.
At the end of the hopping event, particles at sites 1 and 3 (2 and 4)
move on to sites 2 and 4 (1 and 3).
Thus, a system with size $2L$ (let us call it system $A$) 
can be considered two independent systems with size $L$
(call it system $B$), if we interpret the hopping events to the left 
in the system $A$ as a staying event in the system $B$.
Since the system with $p_L=0$ has a bias effect in diffusion
except the pathological case of $p_R = 1$,
the GI for the SCPCPD is in a sense predictable.

As a final remark, we would like to mention how the DPCPD behavior 
can be observed in the BPCPD model. As explained before, the bias
applied to all particles has no effect. 
As was done for the SCPCPD in Ref. \cite{PhP05a}, if different
bias is applied to a particle at singly occupied sites and
a particle at multiply occupied sites, the DPCPD behavior 
such as mean-field-like exponents, logarithmic corrections, {\it etc.}, was
observed (not shown here). 
This unusual bias cannot be included  in the action like
Eq. (\ref{Eq:pcpd_action}) in a simple way, so
this DPCPD behavior is not contradictory to the GIstrong of the BPCPD.

\section{\label{Sec:sum} Summary}
To summarize, 
an efficient algorithm is proposed to simulate the general bosonic
reaction-diffusion systems 
and applies to the single species annihilation model and the bosonic
variant of the pair contact process with diffusion.
For the single species annihilation model, 
renormalization group predictions 
are confirmed numerically.  The BPCPD model is found
to belong to the PCPD universality class and maintains the Galilean
invariance {\it in the strong sense}. Due to the lack of
the analytical predictions for the PCPD, only the comparison of our results
to published simulation results are possible. 

\begin{acknowledgments}
The author acknowledges L. Anton for giving a motivation to make
the algorithm. He also thanks H. Park for helpful discussions about the 
SCPCPD and critical reading of the manuscript.
\end{acknowledgments}
\appendix*
\section{Derivation of E\lowercase{q}. (\ref{Eq:biased})}
From the path integral formalism for RD systems of hard core particles 
introduced in Ref. \cite{PjP05},
Eq. (\ref{Eq:biased}) will be derived in this appendix. 
Since the master equation is linear and the formalism in \cite{PjP05} does not
mix different dynamics, it is enough to consider the diffusion of hard core 
particles. 
For more detailed accounts, see Ref. \cite{PjP05}.

In general, the master equation becomes the imaginary time Schr\"odinger 
equation with (in general non-Hermitian) {\it Hamiltonian} $\hat H$ such that
\begin{equation}
\frac{\partial}{\partial t} | P ; t \rangle
= - \hat H | P;t\rangle,
\end{equation}
where $| P ; t \rangle = \sum_{\{\rho \}} P(\{\rho\}, t)
| \{\rho \}\rangle$
and $| \{ \rho \} \rangle = \prod_\x | \rho_x \rangle$ with 
$\rho_x$ taking either 1 (occupied) or 0 (vacant). 
To write down the {\it Hamiltonian},
introduced are the creation and annihilation operators for hard core particles 
in single species models which satisfy the
following commutation relations:
\begin{equation}
\begin{aligned}
\{ \hat a_\x^\dag, \hat a_\x \} &= 1,\quad
\{ \hat a_\x, \hat a_\x \} = 
\{ \hat a_\x^\dag, \hat a_\x^\dag \} = 0,\\
[ \hat a_\x, \hat a_{\x'} ] &= [\hat a_\x^\dag , \hat a_{\x'}^\dag ] = 0.
\end{aligned}
\end{equation}
Actually, these operators are nothing but the Pauli matrices.
Using creation and annihilation operators,
terms appearing in the {\it Hamiltonian}  
due to diffusion of hard core particles in the single species RD systems
can be written as $\hat H_D = \sum_\x \hat H_\x$ with
\begin{equation}
\begin{aligned}
\hat H_\x =& \sum_{i=1}^d \Bigl [\left (D + \delta_{i,\|}\frac{v}{2}\right )
( \hat n_\x \hat v_{\x + \e_i}  - \hat a_\x \hat a_{\x + 
\e_i}^\dag )\\
&+\left (D - \delta_{i,\|} \frac{v}{2}\right )
( \hat n_\x  \hat v_{\x - \e_i}  - \hat a_\x \hat a_{\x - 
\e_i}^\dag ) \Bigr ],
\end{aligned}
\end{equation}
where $\hat n_\x = \hat a_\x^\dag \hat a_\x$ is the
number operator, $\hat v_\x = 1 - \hat n_\x$, $\e_i$ is the unit vector along
$i$ direction, and hopping is biased along the $\|$ direction.

The differential equation of the generating function
$F$ which is defined as 
\begin{equation}
F(\{ \bar \varphi \} ; t )
\equiv \sum_{\{\rho \} } \left (\prod_\x {\bar \varphi_\x}^{\rho_\x}
\right ) P(\{\rho \} ; t)
= \langle \{ \bar \varphi \} | P ; t \rangle,
\label{Eq:Gen}
\end{equation}
where 
\begin{equation}
\langle \{ \bar \varphi \} | 
 \equiv \prod_\x \left ( \langle 0 |_\x + 
\bar \varphi \langle 1 |_\x \right ),
\end{equation}
takes the form
\begin{equation}
\frac{\partial}{\partial t} F = - \langle \{ \bar \varphi \} |
\hat H | P ; t \rangle.
\end{equation}
The generating function \eqref{Eq:Gen} corresponds to
Eq. (15) of Ref. \cite{PjP05} with the prescription (18a) in Ref. \cite{PjP05}.
Since
\begin{eqnarray}
\langle \{ \bar \varphi \} | \hat a_\x^\dag &=& \bar \varphi_\x 
( 1 - \bar\varphi_\x \hat \varphi_\x ) \langle \{ \bar \varphi \} |,\\
\langle \{ \bar \varphi \} | \hat a_\x&=& \hat \varphi_\x 
\langle \{ \bar \varphi \} |,\\
\langle \{ \bar \varphi \} | \hat n_\x&=& \bar \varphi_\x \hat \varphi_\x  
\langle \{ \bar \varphi \} |,\\
\langle \{ \bar \varphi \} | \hat v_\x&=& (1 - \bar \varphi_\x  \hat \varphi_\x)
\langle \{ \bar \varphi \} |,
\end{eqnarray}
where $\hat \varphi_\x = \partial/\partial \bar \varphi_\x$,
one can find the partial differential equations for the generating function 
such that
\begin{equation}
\frac{\partial}{\partial t} F = - \hat {\cal L} 
(\{ \bar \varphi \}, \{ \hat \varphi\} )F,
\label{Eq:Gen_f}
\end{equation}
with normal ordered evolution operator $\hat {\cal L}$ which reads
\begin{equation}
\begin{aligned}
&\hat {\cal L}
(\{ \bar \varphi \}, \{ \hat \varphi\} ) = \sum_\x \Bigg \{
v \left [ \bar \varphi_\x \partial_\| \bar \varphi_\x - 
\bar \varphi_\x^2 \hat \varphi_\x \partial_\| \hat \varphi_\x\right ]
\\
&+
D \left [ -  \bar \varphi_\x \nabla_\x^2 \hat
\varphi_x + \sum_i (\bar \varphi_\x - \bar \varphi_{\x+\e_i})^2 \hat \varphi_\x 
\hat \varphi_{\x + \e_i}\right ] \Bigg \}\\& + (\textrm{terms due to
reactions})
,
\end{aligned}
\end{equation}
where $\nabla^2_\x$ is the lattice Laplacian defined as
$\nabla^2_\x f(\x) = \sum_{i=1}^d ( f(\x + \e_i)+f(\x - \e_i)- 2f(\x))$,
and $\partial_\|$ is the lattice gradient along the $\|$ direction
defined as $\partial_\| f(\x) = (f(\x + \e_\| ) - f(\x - \e_\|))/2$.
Since Eq. (\ref{Eq:Gen_f}) is a linear equation, we can write down the 
path integral solution with the action \cite{PjP05}
\begin{equation}
S = \int dt  \left [ \bar \phi \partial_t\phi + {\cal L}
(\{ \bar \phi \}, \{ \phi \} ) \right ],
\end{equation}
which completes the derivation of Eq. (\ref{Eq:biased}).


\begin{thebibliography}{99}
\bibitem{comment} There is a numerical integration study of the equivalent 
(complex) Langevin equation in Ref. \cite{BAF97}, 
but the results only confirm the decay exponent.
\bibitem{KCcomment} The observation of the GI for the SCPCPD in 
Ref. \cite{PhP05a} should be understood in the weak sense.
\end{thebibliography}
\end{document}